\documentclass{article} 
\usepackage{graphicx}        
\begin{document}
\centerline{\bf Review of haplotype complementarity under mutational pressure}
\bigskip

Dietrich Stauffer* and Stanis{\l}aw Cebrat
\bigskip

Department of Genomics, 
Wroc{\l}aw University, ul. Przybyszewskiego 63/77, 51-148 Wroc{\l}aw, Poland
\medskip

* visiting from Institute for Theoretical Physics, Cologne University,
D-50923 K\"oln, Euroland
\bigskip

\section{Introduction}

This review deals with the concept of complementarity for diploid genomes,
which was presumably first found in computer simulations \cite{zawierta} 
similar to the old bit-string model of biological ageing \cite{penna,books}. 
We hope that this review (partially taken from \cite{sexschneider}) will
encourage experimental biologists to look for such effects in reality. 
Presumably this complementarity is more likely found in small populations
with low recombination rates during sexual reproduction, and mostly recessive
instead of dominant mutations (our mutations are regarded as detrimental,
causing hereditary diseases).

Usually Darwinian selection is thought to lower the number of detrimental
mutations, but due to copying errors and other reasons
new inheritable mutations appear. Thus the
average number of mutations fluctuates about some low fraction of the total
number of alleles. This evolution we call ``purification''. However, for sexual 
reproduction of diploid organisms, another strategy is possible if (nearly) all 
mutations are recessive; we call this alternative ``complementation''. 

Let us take a simple model of only eight genes; the wild type or functional 
allele is denoted by 0, and the deleterious one by 1. The diploid genome then consists 
of two sequences of 0 and 1, which are called bit-strings. An example may be

\quad \quad  0 0 1 0 0 1 0 0 

\quad \quad  0 0 1 0 1 0 0 0

\noindent
which means that only one locus, the third one, is unfunctional, and all others 
are functional. Only if both alleles are mutated at the corresponding locus the 
mutations affect the phenotype. Thus genes 5 and 6 may be detrimental
in future generations but not for this individual. This example is one of
purification, since only one quarter of the bits are set to 1, and three 
quarters are in the wild form of 0.

An alternative example, for complementation instead of purification, is 

\quad \quad  1 1 0 1 0 1 0 0 

\quad \quad  0 0 1 0 1 0 1 1

\noindent
where now half of the alleles are in the unfunctional state of 1, and half are
still wild (0). Nevertheless, this individual does not feel any 
of these mutations since at no locus both alleles are set to 1. Here we have
complete complementarity. In general, one can change from one extreme to the
other by measuring the heterozygosity, which is the fraction of genetic loci
carrying different alleles in the two bit-strings. This heterozygosity is 
2/8 = 0.25 in the first example and 8/8 = 1.0 in the second example. Thus
purification produces a heterozygosity close to zero, and complementation a 
heterozygosity close to one. The Hamming distance is the number of loci which
are different in the two bit-strings and thus is the heterozygosity multiplied
with the number of investigated loci. (See \cite{eight} for a polymorphic
generalisation to eight instead of only one bit per allele.)

We are not aware that this seemingly trivial possibility of complementarity 
was found earlier than \cite{zawierta}; in hindsight one might interpret 
Fig.1 in \cite{lask} as indicating an evolution towards complementarity.
Now we bring some examples where it was seen in recent computer simulations 
\cite{bonkowska,pekalski,pmco,gamete,zawierta2,kowalski,waga,mackiewicz}.

\begin{figure}[hbt]
\begin{center}
\includegraphics[angle=-90,scale=0.45]{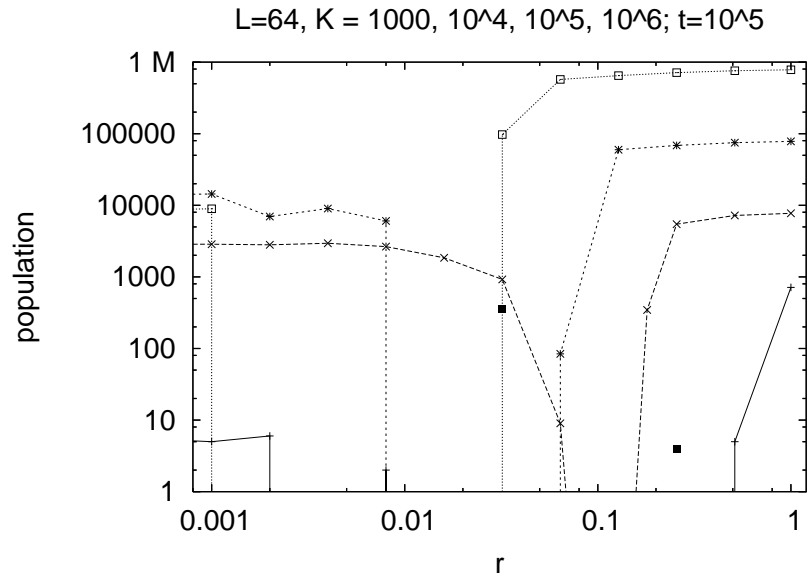}
\end{center}
\caption{Sexual Penna model: populations versus recombination (= crossover) 
rate $r$ for various values of the carrying capacity $K = 10^3 \, (+), \, 
10^4 \, (\times), \, 10^5 \, (*),\, 10^6$ (squares). The gap at 
intermediate $r$ shifts to smaller $r$ for increasing $K$. To the left 
of the gap we have complementarity with many mutated bits; to the right 
we find Darwinian purification selection with much less bits mutated.
}
\end{figure}

\begin{figure}[hbt]
\begin{center}
\includegraphics[angle=-90,scale=0.45]{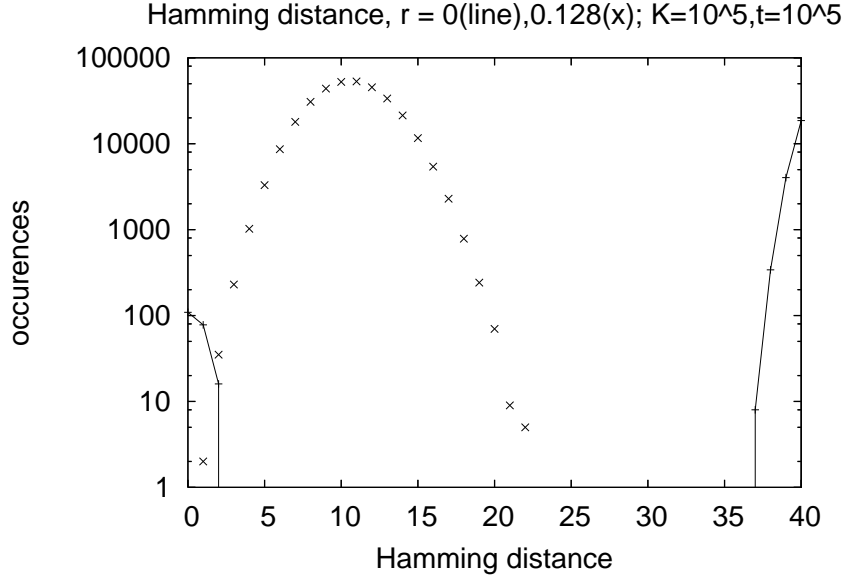}
\end{center}
\caption{Sexual Penna model: Distribution of Hamming distances after
purification (central peak) and
complementation (peaks at the left and right boundaries) for $L=64$ when only
the first 40 bits are compared \cite{bonkowska}. 
}
\end{figure}

\begin{figure}[hbt]
\begin{center}
\includegraphics[angle=-90,scale=0.45]{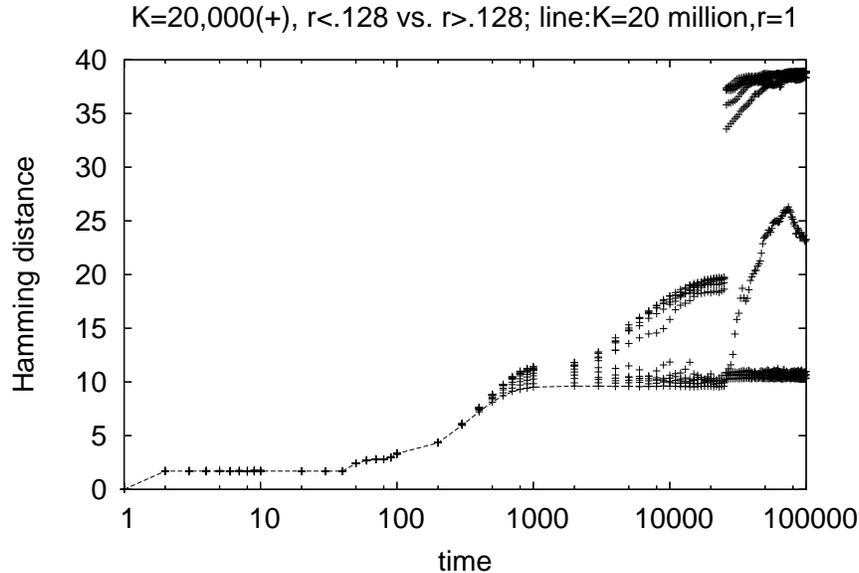}
\end{center}
\caption{Illustration of complementation and gamete recognition for sexual Penna
model at $r = 0$, 0.001, 0.002, 0.004 ... 0.512, 1 from top to bottom. For
small recombination rates $0 \le r \le 0.016$, the average Hamming distances 
approach 20, and then move close to 40 after gamete recognition is switched
on at $t = 25,000$: Complementarity with about half of the 40 bits mutated.
For $r = 0.032$ and 0.064 the population dies out, for $r=0.128$ it does
not know what it wants, and for larger $r$ relatively few bits are mutated:
purification independent of population size.
}
\end{figure}

\section{Mutations}

Mutations can happen due to errors in the genome duplication during cell 
division, or due to external reasons like ionizing radiation.
They may happen in any cell of our body without being transferred to our
children; then they are called somatic and ignored here. Alternatively, they may
appear in the germline cells being transmitted to the offspring by gametes,
and are called inheritable. Most of the mutations make an allele 
unfunctional and thus are deleterious; back mutations to the original ``wild''
state are rare and ignored here. We also do not deal with the rare positive
mutations which transformed the first archeo-bacteria over thousands of million 
years into the present authors. Thus we deal only with deleterious inheritable
mutations. 

At first one may think that life would be better if mutations could be avoided. 
Indeed, sickness, ageing and death may come from such mutations \cite{books}. 
However, if we would live forever, there would me no place for our children,
and biological evolution would not have happened. Indeed, some simulations 
\cite{adapt} in a changing environment showed an optimal mutation rate to
maximise the whole population. Therefore mutational pressure, though bad
for the individual, is not necessarily bad for Nature as a whole. We deal
here with models where ageing or death are caused by this mutational pressure.  
The details of the models are less important than the emergence of 
complementarity in the models.

\section{Some computer models}

The number of bits (genes) in each bit-string (haplotype) is called $L$,
the minimum reproduction age $R$, 
the number of births for each pair after mating $B$, the mutation 
probability per haplotype (i.e. per bit-string) $M$, the probability of
recombination during mating $r$ (for both father and mother; also called the
crossover rate $C$ in the literature). 
The Verhulst factor $N/K$ is the probability to die because of lack of food
or space, where $N$ is the current population and $K$ a ``carrying capacity''.
In the Penna model \cite{penna,books,encybio}, the position of a locus 
corresponds to the
age of an individual, and only mutations at that or at earlier positions 
affect the health of the individual. (New mutations are transmitted to the 
offspring, not to the parent.) $T$ active mutations kill the individual
at that age. Typical values are $L=64, \, R = 5L/8, \, B=4, 
\, M=1, \, 0.001 < r < 1, \, T=3.$ 

For the sexual Penna model, 
Fig. 1 shows the two regimes of low and high 
recombination rates. Each curve has a gap in the middle where the population
dies out, for $r$ near some critical value $r_c$. For low $r$ the population 
survives with the help of the above 
complementarity trick; for high $r$ it survives through purification, i.e. the 
usual Darwinian selection of the fittest with a small number of deleterious mutations.
In the right part of Fig.1, purification happens, and the population is the 
larger the larger $K$ is. In the centre of Fig.1, a gap appears which shifts to
the left with increasing $K$; to the left of the gap, complementarity appears. 
(Increasing the births from $B=2$ to $B=4$ avoids the gap.) Fig.2 shows how 
the equilibrium distribution of Hamming distances looks like after purification 
and for complementation, when only the first $R=40$ of 64 bits are counted.

In such a population nearly all 
individuals have the same bit-strings A and A' in their diploid genome, thus 
producing haploid gametes (ovum and sperm cells) of one type only, either A or 
A'. An A sperm combined with an ovum of gamete type A cannot survive with many 
mutations, since then even homozygous pairs of recessive alleles affect our
health. The same happens with ovum and sperm cell both of type A'. But if one is
of type A and one of type A', the A-A'-zygote can survive even if half of the 
bits (alleles of the genome) are mutated, since there is always a one-bit in A
combined with a zero-bit in A' and thus for recessive mutations the phenotype 
is not affected. Thus high numbers of mutations can be tolerated in this 
strategy. For purification, on the other hand, mutations are rare.
(Warning: Sometimes changes are very slow; for $L=512, \, R=320$ we even had a
case where the population decayed first very slowly, and after 700 million 
iterations very fast to extinction.) 

Somewhat related is gamete recognition \cite{gamete}, where the ovum rejects
those sperm cells for fusion into a diploid zygote whose haploid genome is too 
similar to the haploid genome of the ovum. This effect is beneficial if the
population, due to a low recombination rate, shows complementarity.
If this gamete selection is added to the sexual Penna model then complementarity
survives for higher $r$, the population size to the left of the gap (small
$r$) is strongly enhanced while the populations to the right of the gap 
($r$ closer to unity) barely change. 

Also Fig. 3 illustrates through the Hamming distances this balance between 
complementation at small $r$ and purification at large $r$, separated by 
extinction at intermediate $r$ near $r_c$. For these Hamming distances we take 
into account the first $R=40$ of the 
64 bits. For complementarity without gamete recognition, the whole diploid
population has two bit-strings A and A', each of which with about 20 bits zero 
and 20 bits one. The zygotes thus are of type AA and A'A' with Hamming distances
0 and of type AA' and A'A with Hamming distances 40; the average Hamming 
distance
therefore is 20, as shown in Fig. 3 near $t = 10,000$. The AA and A'A' will die 
out in the next iteration, the AA' und A'A will survive. After 25,000 
iterations, 
gamete recognition is switched on, neither AA nor A'A' is allowed to form a 
zygote, and the Hamming distances approach 40, as shown in the interval
$26,000 \le t \le 100,000$. For large $r$ and purification, the number of 
mutated bits and thus the Hamming distance is much smaller, and the latter 
shows only a small jump from 9.6 to 10.3 (independent of $K$)
when gamete recognition is switched on. (If one of the 64 mutations is made 
dominant, not much changes, but nine dominant mutations lead to catastrophe
\cite{gamete}.)

\begin{figure}[hbt]
\begin{center}
\includegraphics[angle=-90,scale=0.31]{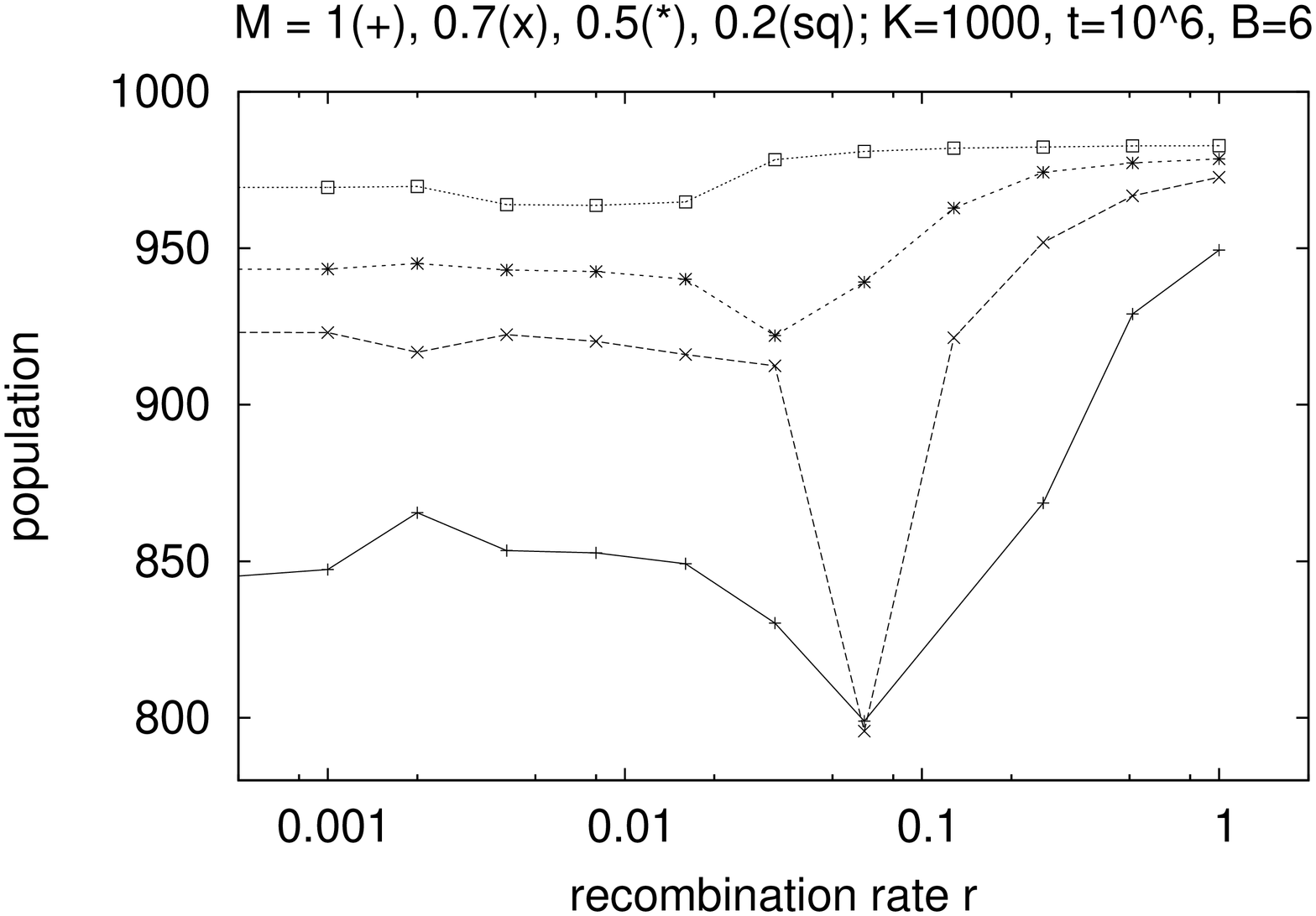}
\includegraphics[angle=-90,scale=0.31]{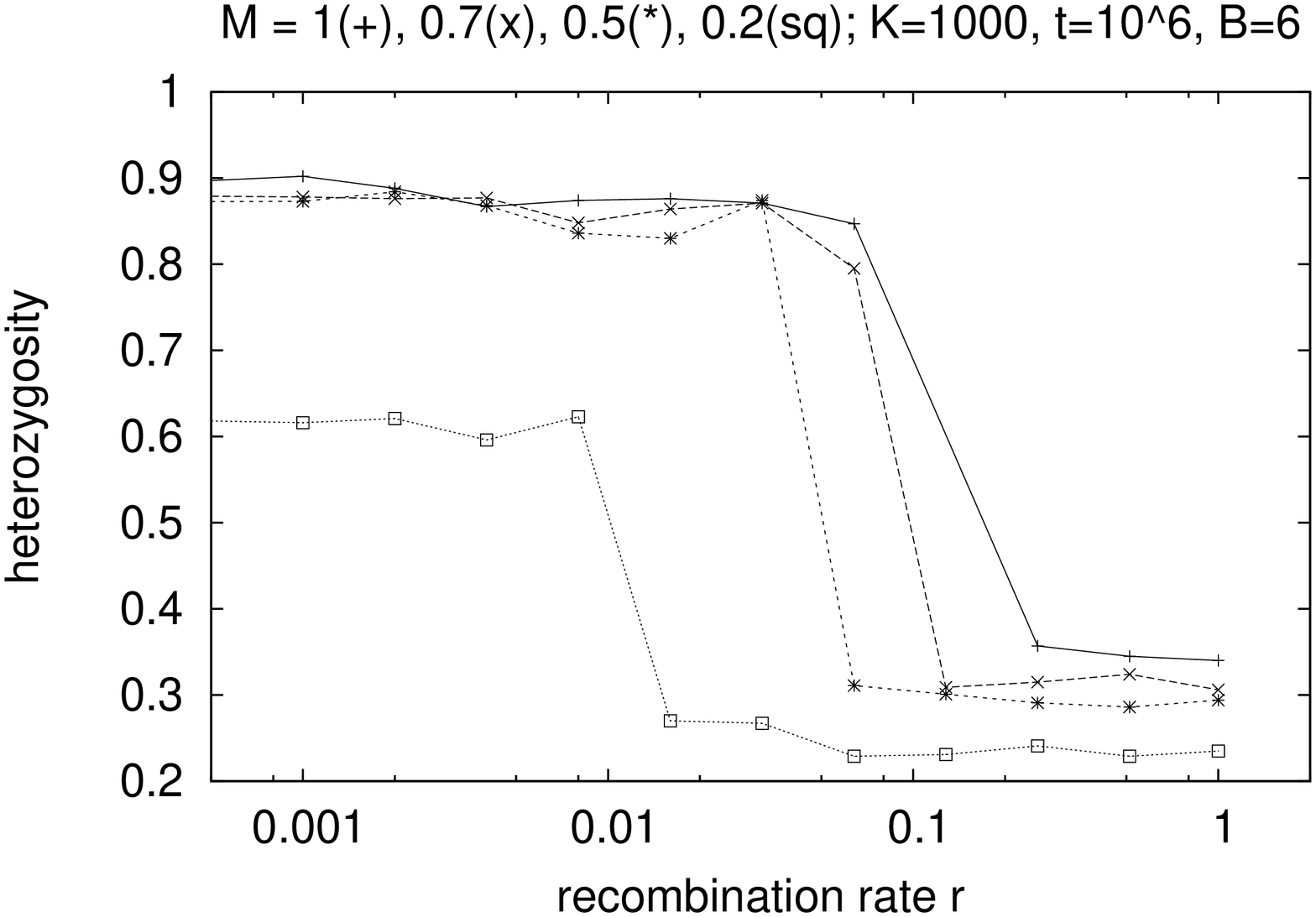}
\end{center}
\caption{Mutation pressure. a: Higher mutation rates lead to a reduction of
the population size and possibly extinction. $L=64, \, R=40, \, B=6, \, T=3,
\, K=1000$. (Similar to Fig.6 in \cite{gamete}.) b: In the left part the 
heterozygosity is higher than in the right part for all these curves. 
}
\end{figure}

\begin{figure}[hbt]
\begin{center}
\includegraphics[angle=-90,scale=0.45]{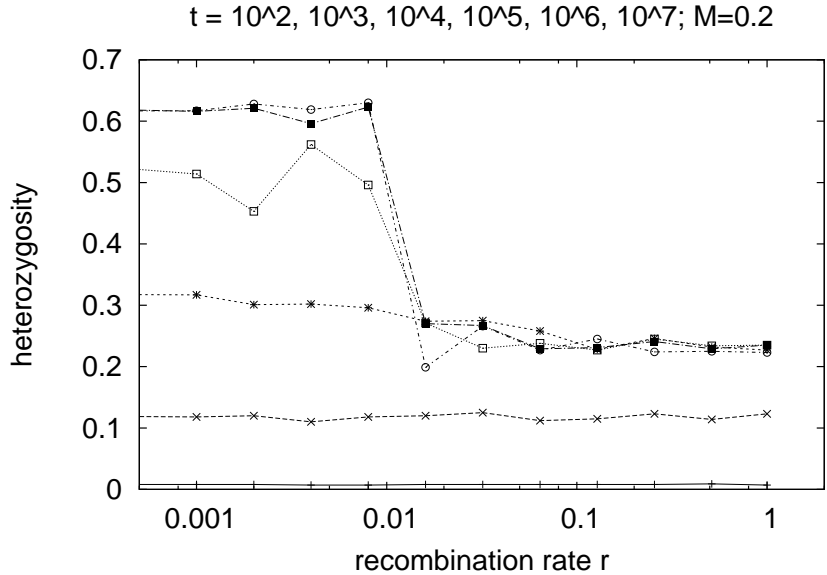}
\end{center}
\caption{Time dependence for the $M=0.2$ curve of the previous figure. For
$M=1$ equilibrium is reached somewhat faster, and $r_c$ is about ten times
higher.
}
\end{figure}

Mutations are needed to drive evolution, but also endanger the survival.
Fig.4 shows the effect of this mutational pressure: For $M=2$ mutations
per generation and haplotype, the population died out while for $M=1$ 
extinction was avoided. For mutation pressure below $M=1$ the population 
reaches nearly its maximum $K$. Fig. 5 shows the slow
emergence of the separation of complementation (small $r$) and purification
(large $r$); here the simulation time $t$ measured in updates per individual
varies from hundred (bottom curve, +) to ten million (top curve, o).

\section{The role of inbreeding}

As has been shown in Figs. 1-3, reproduction success depends on the interplay
between the intragenomic recombination rate (crossover frequency) and the
size of population. Below a specific crossover rate populations prefer to
complement haplotypes instead of to intensively eliminate defective alleles. In
Fig.6 we show how this critical crossover rate depends on the population size 
\cite{zawierta2}, where $r_c$ is defined as the crossing probability at which
the number of mutations goes down drastically. In the range of two
decades there is a power law relation. Nevertheless, the data shown in the
plot were obtained in simulations of panmictic populations. In such
populations females look for and choose randomly a sexual partner from the
whole population. In Nature the process of choosing the partner is usually
nonrandom and, what is more important, it is spatially restricted.
Individuals are looking for partners in their neighbourhood. Thus, the effect
of the population size should be considered as an effect of the inbreeding,
rather. Inbreeding (coefficient) is a measure of genetic relations between
individuals. If the individuals live in small ``inbreeding'' groups, then the
inbreeding coefficient is high and there is a high probability that the sexual
partners share some fragments of the same ancestral genome.

\begin{figure}[hbt]
\begin{center}
\includegraphics[angle=-90,scale=0.45]{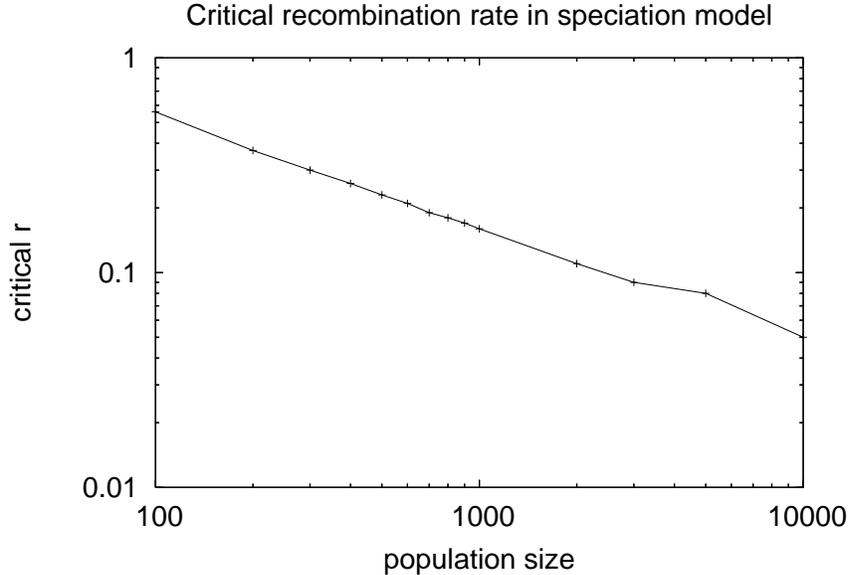}
\end{center}
\caption{Log-log plot of critical recombination rate $r$ versus population size 
\cite{zawierta2}. For $r$ below this value, complementary instead of purified
haplotypes (bit-strings) are preferred. 
}
\end{figure}

To study the effect of inbreeding, the simulation of evolution was performed
on lattices; see \cite{bonkowska} for inbreeding without lattices, by dividing
the population into groups. On lattices the level 
of inbreeding was set by declaring the maximum
distance where individuals can look for partners and where they can place
their offspring \cite{zawierta}. The simulations were performed on a square
lattice $1000 \times 1000$. (Indeed, if the lattice size varies with a fixed 
size of the neighbourhood, $r_c$ barely changes \cite{zawierta2}). 
If the above distances within which partners are
searched were set to 5, the critical
crossover rate $r_c$ was around 0.2. Populations evolving under lower
recombination rate or shorter distances prefer the strategy of complementing
the haplotypes while under higher recombination rate or longer distances
they choose the strategy of purifying selection. Nevertheless, there are
very important consequences of such a kind of choice. The complementarity
evolves locally and remote subpopulations on the same lattice can have
different distributions of defective alleles in their haplotypes. Using some
tricks with coloring the individuals according to their genomes' structure
it has been shown that the lattice is occupied by individuals with different
genotypes but they are clustered. Individuals with the same genotypes
occupy the same territory (see http://www.smorfland.uni.wroc.pl/sympatry/
for some examples of simulations under different conditions). 

(Further studies have shown that for sympatric speciation 
only the central part of the genome is responsible. The lateral
part of the genome is much more polymorphic and decides on biodiversity,
rather than speciation. That is why the Hamming distances between homologous
haplotypes inside species are noticeable. These simulations show that
sympatric speciation is possible and there is no need for physical,
geographical or even biological barriers for the new species to emerge inside
the population of the older one.)

\begin{figure}
\begin{center}
\includegraphics[angle=-90,scale=0.45]{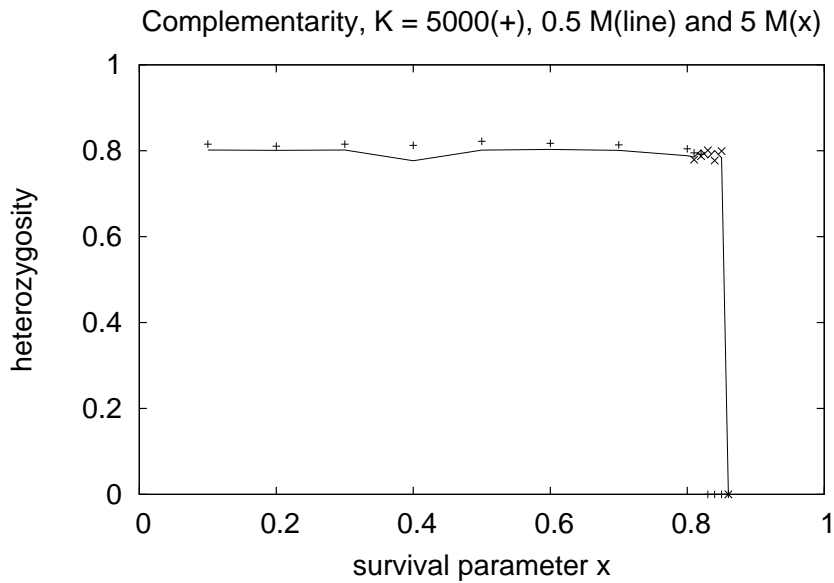}
\end{center}
\caption{Jump as a function of $x$ at crossover probability $r=0.01$.
}
\end{figure} 

However, complementarity is not {\em always} a strong function of the population
size. The results of \cite{cebrat} are unclear; and in the simple model of 
\cite{stauffer}, Fig.7 shows about the same transition from complementation
to purification, when the population size is increased by a factor of thousand. 
We see practically no change whether $K$ is 5000, half a million, or five 
million. (In that simplification of \cite{pmco}, no age structure is involved,
and an individual survives with probability $x^nV$ where $V$ is the usual
Verhulst factor for adults, $n$ the number of deleterious mutations appearing 
in both bit-strings (chromosomes) of the diploid genome, and $x < 1$ determines
the damage made by a single mutation present in both bit-strings. Males and
females are distinguished. New mutations occur at gamete production and are 
transferred only to the baby, not to the parent. (In \cite{stauffer} we did 
not yet search for this complementarity shown now in Fig.7.)

\section{Consequences} 

Now that the reader may have understood the complementarity principle,
what are the consequences of this possible survival strategy via 
complementarity? 
Also, complementarity is an advantage of sexual reproduction
compared to the asexual haploid case where complementarity is impossible.
Now we discuss some further consequences:

{\bf Sympatric Speciation:} If one species splits into two within the same 
environment, this is called sympatric speciation. It is facilitated by
complementarity through the following effect. Originally we have complementary
haplotypes A and A' as discussed in section 3, leading to survivable AA' or A'A 
zygotes even though only about half of the alleles are of the wild type.
Slowly, for part of the 
population A may be change into B, and simultaneously A' into the complement B'
of B. After some time, AB' and A'B zygotes may no longer be survivable, and the
subpopulation with B and B' has become reproductively isolated from that with
A and A' haplotypes. For purification, in contrast, we have only A changing into
B while keeping most alleles in the wild type, thus still keeping AB zygotes
survivable because of the low number of deleterious alleles. In this way,
reproductive isolation and thus speciation is easier for complementarity
than for purification.

{\bf Preferred loci for recombination:}
Computer simulations have shown that a critical parameter for the emergence of
complementarity is the recombination frequency. Human 
genome parameters suggest that the consequences of complementarity should 
be seen at least in some regions of human chromosomes, especially if one 
considers the uneven distribution of recombination events along them. 
There are so-called recombination hot spots observed where recombinations 
happen relatively often, and recombination deserts where recombinations are 
not observed at all. In these deserts complementing clusters of genes 
should appear. Moreover, these regions seem to be non-randomly distributed 
on chromosomes.

It has been noticed that the distribution of accepted recombination events 
in the genomes of simulated populations depends on parameters of 
simulations. If evolution is studied in small effective populations under 
relatively low recombination rates, the central parts of chromosomes 
start to form clusters of genes where recombinations have deleterious 
effect on reproduction potential. Gametes which are produced by 
recombinations in these regions have lower chance to produce the surviving 
zygotes. As a result, the recombination events in gametes which succeeded 
in forming the surviving individuals have a characteristic distribution, 
with higher recombination frequencies in the regions close to the ends of 
chromosomes and lower recombination rate in the 
central part of chromosomes, as observed in reality. 



{\bf Gamete recognition:}
Complementation strategy assumes that two different (complementing) 
sequences of alleles fit to each other producing a better fitted genome. 
If we consider a set of chromosomes with only one pair of complementing 
clusters then it would be more economical to recognize which chromosome 
has an identical cluster and which one has a complementing cluster of genes,
before two gametes fuse to form a zygote. Such systems of recognition or 
probing the information inside another cell are known even in the bacteria 
world - i.e. an entry exclusion system which prevents a bacteria to engage in 
conjugational process if a partner cell already possesses genetic information 
to be transferred \cite{nov,gamete}. It is suggested that in humans the Major 
Histocompatibility Complex (MHC) can play such a role in preselection of 
partners \cite{odor,pekalski}. This complex alone is not enough to guarantee 
the fusion 
of complementing haplotypes. The mechanism should be located at the level of 
gametes and, to be efficient, it should be independent for different pairs 
of chromosomes nestling the complementing clusters of genes.  There is a 
group of genes which could fulfil such a role - Olfactory Receptor genes 
(OR). This is the largest gene family in the human genome composed of almost 
1000 genes and pseudogenes, clustered in many different groups located on 
almost all chromosomes (excluding Y) and at least some of these genes are 
expressed during spermatogenesis \cite{spehr}. If we assume that each of our 
22 pairs of autosomes has complementing clusters of genes, then an ovum 
would have extremely low chance to find a fully complementing sperm cell 
($2^{-22})$. If an ovum could choose such a sperm cell, it should have a pool 
of at least $2^{-22}$ sperm cells. In fact this pool seems about 10 times 
larger.

\begin{figure}
\begin{center}
\includegraphics[angle=-90,scale=0.45]{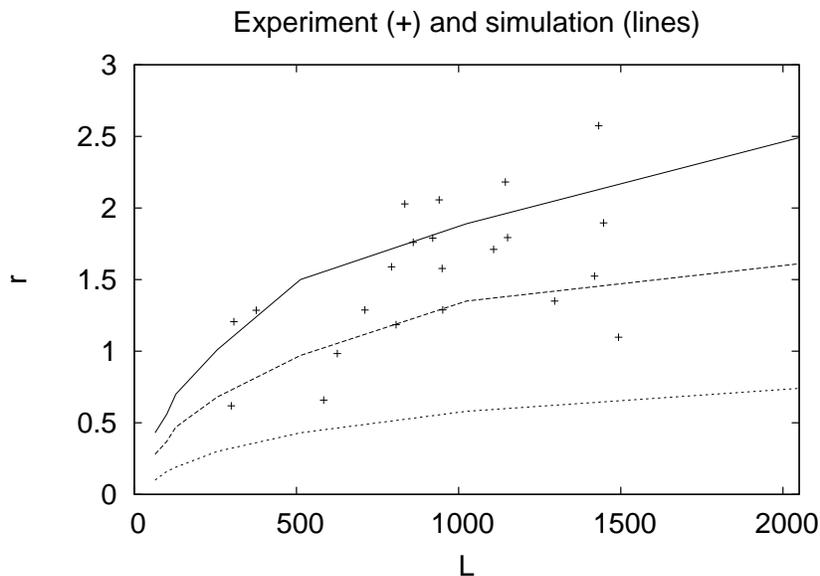}
\end{center}
\caption{
Comparison of simulated critical recombination rate $r_c$ for effective
population size 100, 200, 1000 (from top to bottom) for chromosomes
containing $L$ genes with parameters of human chromosomes. For real
chromosomes (+), data show the average number of crossovers per meiosis
(y-axis) against the number of genes per chromosome (x-axis). Thus human 
genomic data suggest that at least some parts of our genome could evolve 
under complementing regime.
}
\end{figure}

\section{Discussion}

The destruction of complementarity by high crossover rates $r$ is easy to 
understand: The delicate emergence of two complementary bit-strings A and A' 
in the whole population is destroyed for each individual where crossover in 
the middle of the chromosome leads to massive changes in the chromosome 
structure.  The dependence on the 
(effective) size of the population seems more complicated. It is also possible
that $r_c$ depends on the size of the chromosome, or that in one genome some 
chromosomes should complement while others follow purification \cite{waga}.
Fig.8 compares such simulations with reality; the order of magnitude seems 
to be realistic.
Complementarity may also affect the distribution of crossing points along the
bit-strings \cite{kowalski,mackiewicz}. Complementarity requires that whole
sequences of neighbouring genes are transmitted together after recombination;
if for each locus the transmitted allele is selected randomly one would hardly
find nearly complementary haplotypes \cite{mackiewicz}. In any case, the 
details of the models are not important; our important point is that allele 
complementarity is in principle plausible, was found in some computer 
simulations, and should be checked in reality.

\end{document}